\documentclass[twocolumn]{aastex63}

\usepackage{color,comment,multirow,booktabs,upgreek}

\graphicspath{{./}{images/}}

\newcommand{\co}{$^{12}$CO(2$\rightarrow$1)} 
\newcommand{\hi}{\ifmmode{\rm HI}\else{H\/{\sc i}}\fi} 
\newcommand{\sone}{SMC-C1} 
\newcommand{\stwo}{SMC-C2} 
\newcommand{\glon}{\ifmmode{\ell}\else{$\ell$}\fi} 
\newcommand{\glat}{\ifmmode{b}\else{$b$}\fi} 
\newcommand{\vlsr}{\ifmmode{V_\mathrm{LSR}}\else{$V_\mathrm{LSR}$}\fi} 
\newcommand{\vwind}{\ifmmode{V_\mathrm{w}}\else{$V_\mathrm{w}$}\fi} 
\newcommand{\de}{\ifmmode{^\circ}\else{$^\circ$}\fi} 
\newcommand {\kms}{\ifmmode{\rm km \, s^{-1}}\else{$\rm km \, s^{-1}$}\fi} 
\newcommand {\mo}{{\rm M}_\odot}

\newcommand {\moyr}{\,{\rm M_\odot\,\rm yr}^{-1}}

\newcommand{\eqref}[1]{Eq.\ (\ref{#1})}

\newcommand{\alphaco}{\alpha_\mathrm{CO}}
\newcommand{\alphacoun}{\mo \, (\mathrm{K} \, \kms \, \mathrm{pc^2})^{-1}}

\newcommand{\coint}{F_\mathrm{CO}}
\newcommand{\colum}{L_\mathrm{CO}}
\newcommand{\hs}{\hspace{17pt}}

\defcitealias{McG+18}{MCG18}

\shorttitle{Molecular gas in the SMC outflow}
\shortauthors{Di Teodoro et al.}

\begin{document}

\title{Molecular gas in the outflow of the Small Magellanic Cloud}%

\correspondingauthor{E.~M. Di Teodoro}
\email{enrico.diteodoro@anu.edu.au}

\author[0000-0003-4019-0673]{Enrico M. Di Teodoro}
\affiliation{Research School of Astronomy and Astrophysics - The Australian National University, Canberra, ACT, 2611, Australia}

\author[0000-0003-2730-957X]{N. M. McClure-Griffiths}
\affiliation{Research School of Astronomy and Astrophysics - The Australian National University, Canberra, ACT, 2611, Australia}

\author[0000-0002-6637-3315]{C. De Breuck}
\affiliation{European Southern Observatory, Karl-Schwarzschild-Str. 2, 85748 Garching bei München, Germany}

\author{L.\ Armillotta}
\affiliation{Research School of Astronomy and Astrophysics - The Australian National University, Canberra, ACT, 2611, Australia}

\author{N.\ M.\ Pingel}
\affiliation{Research School of Astronomy and Astrophysics - The Australian National University, Canberra, ACT, 2611, Australia}

\author{K.\ E.\ Jameson}
\affiliation{Research School of Astronomy and Astrophysics - The Australian National University, Canberra, ACT, 2611, Australia}

\author{John\ M.\ Dickey}
\affiliation{School of Natural Sciences, University of Tasmania, Hobart TAS, Australia}

\author{M.\ Rubio}
\affiliation{Departamento de Astronom\'{i}a, Universidad de Chile, Casilla 36, Santiago de Chile, Chile}

\author{S.\ Stanimirovi\'{c}}
\affiliation{Department of Astronomy, University of Wisconsin, Madison, WI 53706, USA}

\author{L.\ Staveley-Smith}
\affiliation{International Centre for Radio Astronomy Research (ICRAR), University of Western Australia, Crawley, WA 6009, Australia}
\affiliation{ARC Centre of Excellence for All Sky Astrophysics in 3 Dimensions (ASTRO 3D)}

\begin{abstract}
We report the first evidence of molecular gas in two atomic hydrogen (\hi) clouds associated with gas outflowing from the Small Magellanic Cloud (SMC). 
We used the Atacama Pathfinder Experiment (APEX) to detect and spatially resolve individual clumps of \co\ emission in both clouds. 
CO clumps are compact ($\sim 10$ pc) and dynamically cold (linewidths $\lesssim1 \, \kms$).
Most CO emission appears to be offset from the peaks of the \hi\ emission, some molecular gas lies in regions without a clear \hi\ counterpart.
We estimate a total molecular gas mass of $M_\mathrm{mol}\simeq10^3-10^4\, \mo$ in each cloud and molecular gas fractions up to $30\%$ of the total cold gas mass (molecular + neutral).
Under the assumption that this gas is escaping the galaxy, we calculated a cold gas outflow rate of $\dot{M}_\mathrm{gas}\simeq0.3-1.8 \, \moyr$ and mass loading factors of $\beta \simeq 3 -12$ at a distance larger than 1 kpc. 
These results show that relatively weak starburst-driven winds in dwarf galaxies like the SMC are able to accelerate significant amounts of cold and dense matter and inject it into the surrounding environment.
\end{abstract}

\vspace*{1cm}

\section{Introduction}

Galactic outflows powered by either active galactic nuclei (AGN) or star formation feedback have been observed in many galaxies \citep[e.g.][]{Veilleux+05}. 
Winds have a primary impact in many aspects of galaxy evolution, for example in regulating the efficiency of star formation and in enriching the circum-galactic medium with metals \citep[][for a review]{Zhang18}.
The multiphase nature of gas in outflows, from the hot highly-ionized phase (10$^{6-7}$ K) down to the cold molecular phase ($<100$ K), has been confirmed both by observations \citep[e.g.,][]{Arribas+14,Leroy+15,Martin+16} and by numerical simulations \citep[e.g.,][]{Tanner+16, Kim&Ostriker18,Armillotta+19}.

\begin{figure}[t]
\center
\includegraphics[width=0.47\textwidth]{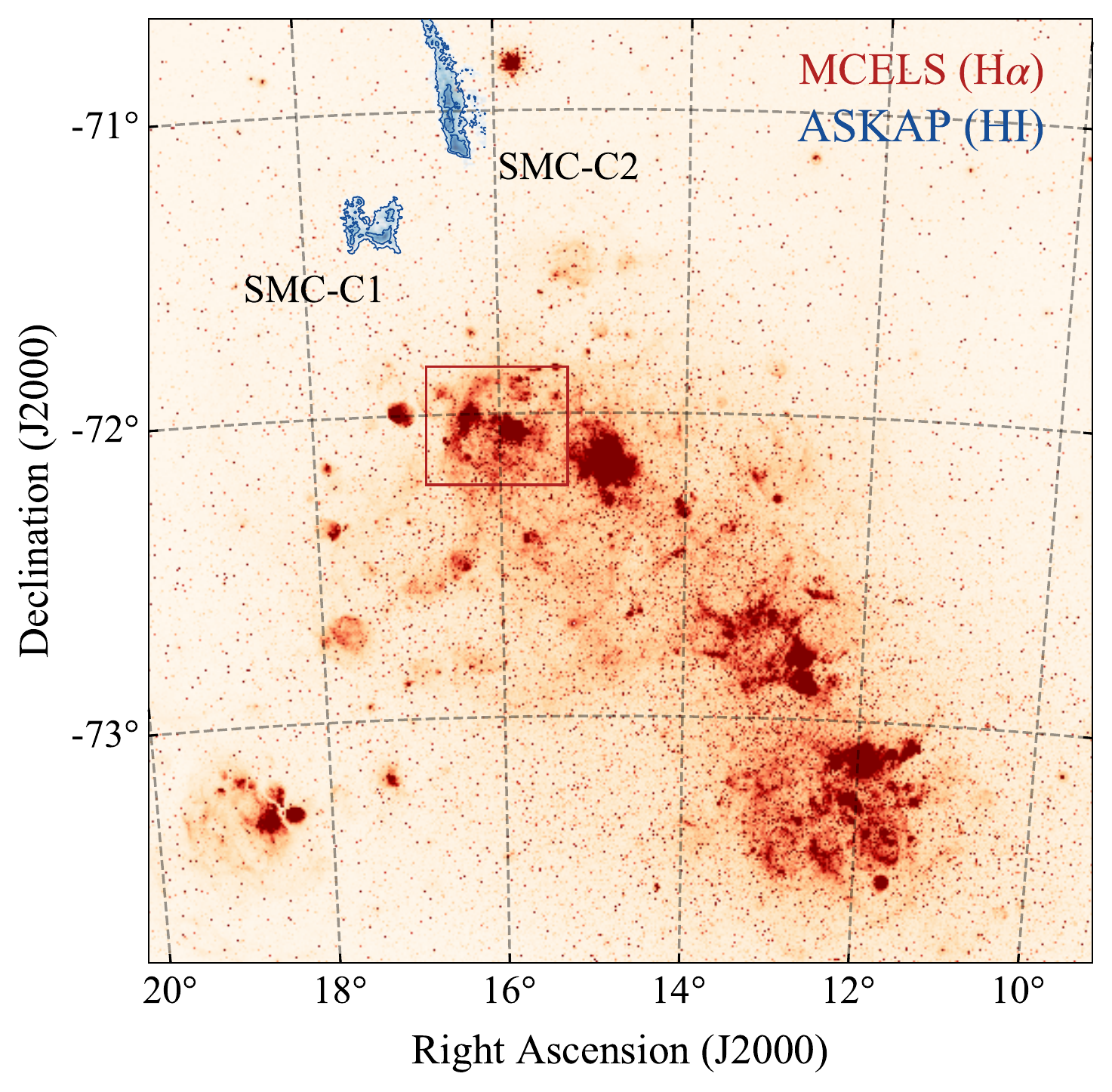}
\caption{The Small Magellanic Cloud (SMC) in H$\upalpha$ emission \citep[MCELS survey,][]{Winkler+15} in red color scale. The two clouds (\sone\ and \stwo) detected in \hi\ emission \citep[ASKAP,][]{McG+18} and targeted in this work are shown in blue color scale.
\hi\ contours (blue) are at $(1.0,2.0)\times10^{19}$ cm$^{-2}$ (\sone) and $(1.5,2.5)\times10^{19}$ cm$^{-2}$ (\stwo).
The red box denote the star-formation regions (NGC 371/395) from where the clouds could have been launched. 
}
\label{fig:largemap}
\end{figure}

Because of their shallow gravitational potential, dwarf galaxies are particularly sensitive to supernovae (SNe) explosions and stellar winds and can easily generate galactic winds.
Simulations of galaxy evolution with stellar feedback predict that dwarf galaxies can expel large amounts of gas on kiloparsec scales and that cool gas ($T<10^4$ K) contains most to half of the mass of the outflow \citep[e.g.][]{Hopkins+12,Muratov+15}. 
Tidal interactions and mergers can cause bursts of star formation that furthermore power the outflow \citep{Hopkins+13}.
Being the closest pair of interacting dwarf galaxies, the Magellanic Clouds constitute unique laboratories to study on parsec scales stellar feedback in action in non-isolated gas-rich dwarf galaxies.

Recently, \citet{McG+18} (MCG18, hereinafter) demonstrated for the first time that a significant amount of neutral gas is outflowing from the main body of the Small Magellanic Cloud (SMC). 
Using high-resolution atomic hydrogen (\hi) emission-line data with the Australian Square Kilometre Array Pathfinder (ASKAP), they detected a population of clouds and filaments consistent with being gas driven out by the intense star-formation regions in the SMC. 
They estimated a \hi\ mass in the outflow of $\sim10^7 \, \mo$, about 3\% of the total atomic gas mass of the galaxy, with an \hi\ outflow rate of $0.2-1\,\moyr$, i.e.\ $2-10$ times larger than the galaxy star formation rate (SFR).
However, the amount of molecular gas entrained in the outflow, which is expected to be significant, is still unconstrained.

\begin{figure*}
\center
\includegraphics[width=1\textwidth]{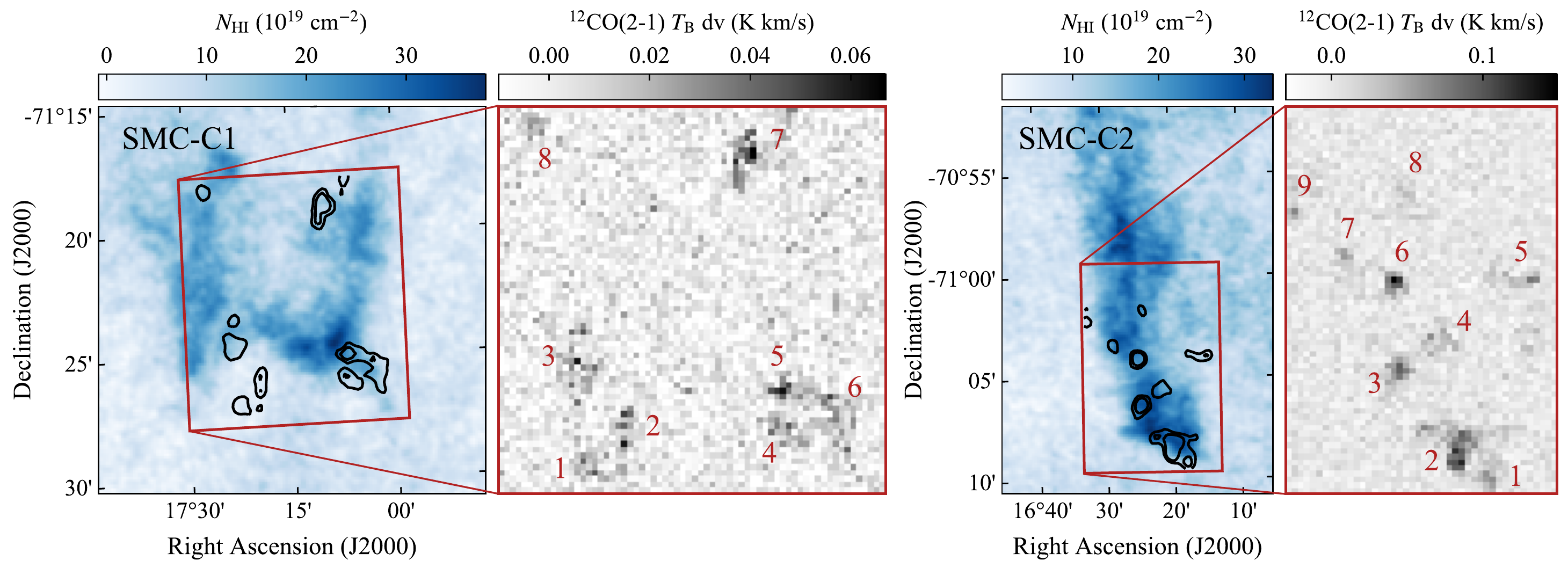}
\caption{\hi\ and CO maps of \sone\ (left panels) and \stwo\ (right panels). 
Column-density maps of the \hi\ emission from ASKAP observations \citep{McG+18}, integrated over the velocity range $90-115 \, \kms$ (\sone) and $120-140\,\kms$ (\stwo), are shown in blue color-scale. 
The fields mapped in the \co\ emission line with APEX are delimited with red boxes.
Greyscale maps represent the integrated intensity of the \co\ emission line. 
We identify eight clumps of molecular gas in \sone\ field and nine in \stwo, labelled in red in the CO maps.
On the \hi\ maps, we overlay the CO contours at levels [0.02, 0.04] K km/s.
}
\label{fig:clouds}
\end{figure*}

In this work, we quantify the contribution of molecular gas to the mass loss budget in the SMC outflow.
We use the Atacama Pathfinder Experiment \citep[APEX,][]{Gusten+06} telescope to detect and study \co\ emission in two outflowing \hi\ clouds.
In the remainder of this Letter, we introduce our new APEX observations in \autoref{sec:data}, we  describe our main findings in \autoref{sec:results} and discuss them in \autoref{sec:discussion}, summarizing in \autoref{sec:conclusions}.
Throughout this Letter, we assume a distance for the SMC of $D = 63\pm3$ kpc, for which 10 arcsec correspond to about 3 pc.

\section{Observations}
\label{sec:data}

We targeted two \hi\ clouds (\sone\ and \stwo\ thereafter) identified by \citetalias{McG+18} and believed to be material expelled because of star formation feedback. 
These clouds are amongst the most prominent and highest-density features with anomalous kinematics in the SMC and they show some intriguing velocity gradients across their structures, suggestive of acceleration.
\autoref{fig:largemap} displays a large-scale H$\upalpha$ map of the SMC (red color scale) from the Magellanic Cloud Emission-line Survey \citep[MCELS,][]{Winkler+15}, showing several regions of intense star formation. 
\hi\ column-density maps of our targets are overplotted in \autoref{fig:largemap} and shown in detail in \autoref{fig:clouds} (blue color scale). 
For this work, we mapped the \co\ line with APEX in the regions outlined by the red boxes in \autoref{fig:clouds}. Details on the observations and data reduction are given in \autoref{app}.

\section{Results}
\label{sec:results}

\subsection{Morpho-kinematics of the CO emission}

Our APEX data show some weak but unequivocal \co\ emission throughout the two fields in correspondence of the velocity range of the \hi\ emission, i.e.\ local standard of rest (LSR) velocity $\vlsr = 90-115 \, \kms$ for \sone\ and $\vlsr = 120-140 \, \kms$ for \stwo.
To identify regions of genuine emission, we used the 3D source finder implemented in the $^\mathrm{3D}$\textsc{Barolo} code \citep{DiTeodoro&Fraternali15}. 
In short, the source finder smooths the data to a lower spatial resolution ($50''$ in our case) to improve the signal-to-noise ratio, and reconstructs the sources by merging regions with flux higher than a given threshold ($3\times\mathrm{rms}$) that are close in both the spatial and the spectral domains.
We detected eight main knots or clumps of CO emission in the \sone\ field and nine main knots in the \stwo\ field.
A few other regions of low-significance emission were not included in our further analysis. 

\autoref{fig:clouds} shows \co\ maps (grey color scale) integrated over the velocity channels showing emission. 
The detected CO clumps are labelled in red. 
We overlay the contours of the CO emission on the \hi\ map of \autoref{fig:clouds}. 
In general, in both fields, the peaks of the CO emission are offset from the regions with the highest \hi\ column densities. 
In \stwo, most of the molecular gas is well aligned and shares the kinematics with the neutral gas filament visible in \hi.
The CO emission in \sone\ has lower signal-to-noise and the association with the \hi\ emission is not as striking as in \stwo: for example, clumps 1, 2 and 7 are located in regions with no or very low column-density neutral gas; clump 4 does not match the kinematics of the adjacent \hi\ gas and has a velocity about 10 $\kms$ lower than the \hi\ emission. 
Typical sizes of the CO clumps are of the order of 10 pc.

\begin{figure*}
\center
\includegraphics[width=1\textwidth]{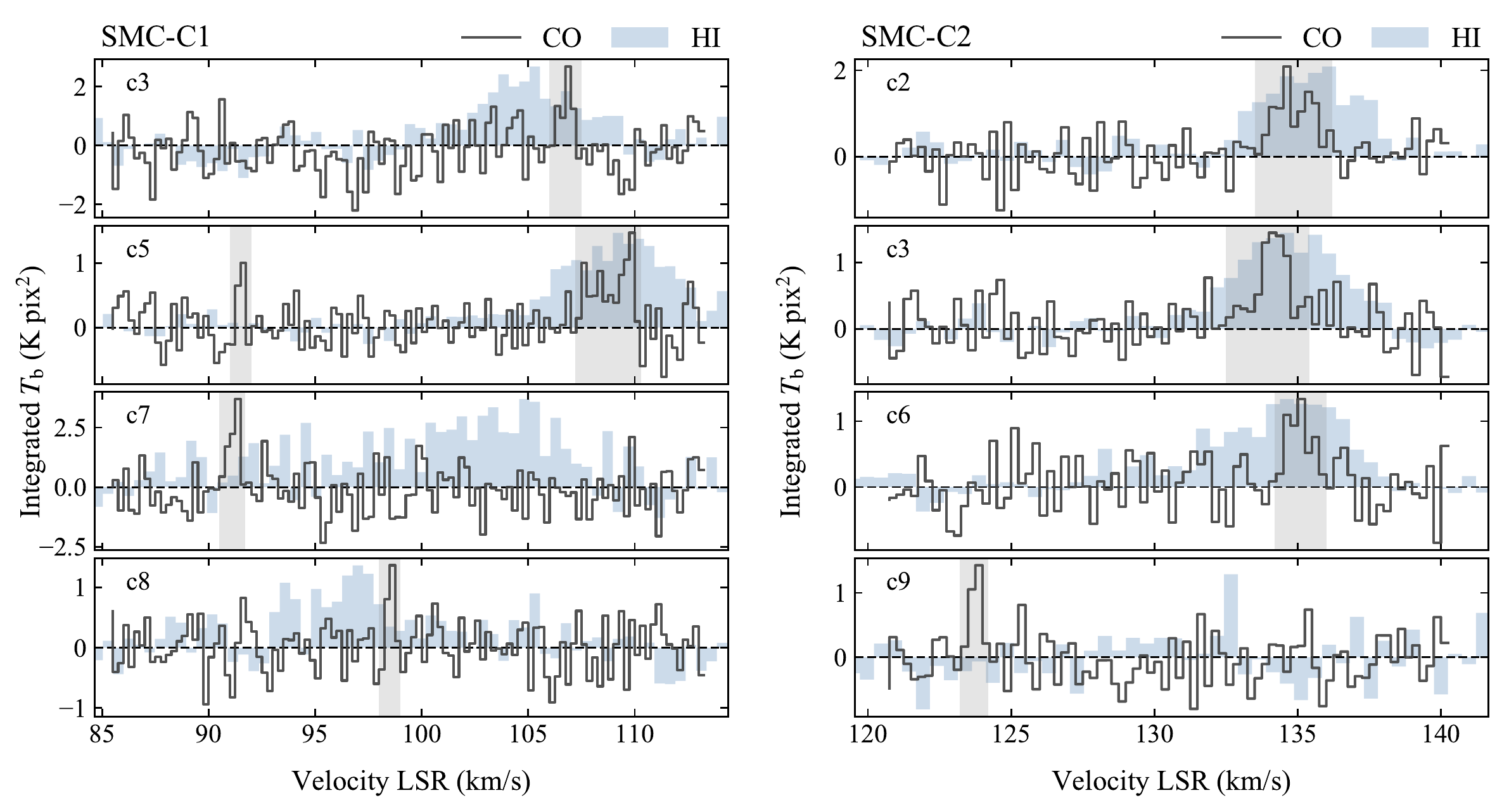}
\caption{
Spectra of the main clumps of \co\ emission (black line) detected with APEX for \sone\ (left panels) and \stwo\ (right panels). 
Spectra are integrated over a $0.7'$ circular aperture. 
The velocity range of each detection is highlighted in grey.
The narrow emission line at $V_\mathrm{LSR}\simeq91\,\kms$ visible in the \sone's clump 5 spectrum is contamination from the nearby, spectrally decoupled, clump 4.
Light-blue histograms denote \hi\ spectra integrated on the same aperture and normalized to the CO peak for a easier visualization. 
}
\label{fig:spectra}
\end{figure*}

In \autoref{fig:spectra}, we plot spectra for some representative CO knots in \sone\ (left panels, clumps 3, 5, 7 and 8) and in \stwo\ (right panels, clumps 2, 3, 6 and 9). 
Spectra are integrated over circular regions with radius of $0.7'$ centered on the emission peak.
CO emission lines are clearly detected. 
The spectral range corresponding to each detection reconstructed by the source finder is highlighted in grey.
We also plot \hi\ spectra (blue histograms) in the same regions extracted from new, high spectral resolution (0.5 $\kms$), ASKAP \hi\ data (Pingel et al., in prep).
Some of the stronger and more extended knots, like clump 5 in \sone\ and clump 2 in \stwo, show some velocity gradient across their structure, which results in a relatively broad emission over 1-2 $\kms$ having multiple kinematical components.
These knots have velocities in agreement with the broader \hi\ emission.
Fainter knots (e.g., clump 8 in \sone\ or 9 in \stwo) have very narrow, single-component emission, with FWHM line-broadening lower than 0.5 $\kms$ and can be offset in velocity from the \hi\ emission.
In general, all CO clumps have modest line-widths, suggesting that turbulence does not play an important role in shaping the molecular gas in the wind.
 
The main properties of the CO knots are listed in \autoref{tab:prop}. 
Central velocities (column 2), line-widths (column 3) and integrated flux densities (column 4) are measured through a Gaussian fit of the spectrum integrated over the entire 3D detection. 
The radius (column 6) of a clump is calculated as $R_\mathrm{c}=D\tan(\sqrt{\Omega_\mathrm{s}/\pi})$, where $\Omega_\mathrm{s}$ is the area covered by all detected pixels in the integrated intensity map.

\begin{table*}
\centering
\caption{Properties of the CO clumps in \sone\ (a) and \stwo\ (b). Columns: (1) clump number; (2)-(3) central velocity (LSR) and FWHM linewidth of the integrated CO line; (4) Integrated flux density; (5) CO luminosity; (6) Clump size in radius $R_\mathrm{c}=D\tan(\sqrt{\Omega_\mathrm{s}/\pi})$ (see text); (7) Molecular mass using a $\alphaco=0.12 \, \alphacoun$, i.e.\ optically-thin regime for the \co\ transition; (8) Molecular mass using a $\alphaco=2 \, \alphacoun$. 
Errors are propagated from the uncertainties of the Gaussian fit to the integrated spectrum of each knot.}
\label{tab:prop}
\begin{tabular}{lccccccc}
\multicolumn{8}{c}{(a) \sone}  \\
\noalign{\vspace{5pt}}\hline\hline\noalign{\vspace{5pt}}
\multirow{2}{*}{\# \hspace{5pt}} &
\multirow{2}{*}{\hs $\vlsr$ \hs} & 
\multirow{2}{*}{\hs FWHM \hs} & 
\multirow{2}{*}{\hspace{20pt} $F_\mathrm{CO}$\hspace{20pt}} & 
\multirow{2}{*}{\hspace{25pt} $L_\mathrm{CO}$\hspace{25pt}} & 
\multirow{2}{*}{\hs $R_\mathrm{c}$\hs} & 
\multicolumn{2}{c}{\hspace{30pt}$M_\mathrm{mol}$ ($\mo$) \hspace{30pt} } \\ \cmidrule{7-8}
    & (km/s) & (km/s) & (K km/s) & (K km/s pc$^2$) & (pc) & $\alpha_\mathrm{thin}$ & $\alpha_\mathrm{thick}$ \\
\noalign{\vspace{5pt}}\hline\noalign{\vspace{5pt}}
1 & $91.89\pm0.12$ & $1.11\pm0.18$ & $1.12\pm0.36$ & $326.4\pm124.2$ & $10\pm2$ & $39\pm14$ & $652\pm248$ \\ 
2 & $93.12\pm0.14$ & $0.28\pm0.19$ & $0.46\pm0.32$ & $150.1\pm111.9$ & $11\pm2$ & $18\pm13$ & $300\pm223$ \\ 
3 & $106.98\pm0.18$ & $0.48\pm0.18$ & $1.34\pm0.52$ & $645.9\pm299.3$ & $13\pm2$ & $77\pm35$ & $1291\pm598$ \\ 
4 & $91.56\pm0.11$ & $0.40\pm0.15$ & $1.09\pm0.42$ & $373.6\pm170.3$ & $11\pm2$ & $44\pm20$ & $747\pm340$ \\ 
5 & $108.75\pm0.15$ & $1.52\pm0.16$ & $2.36\pm0.43$ & $1186.4\pm294.5$ & $13\pm2$ & $142\pm35$ & $2372\pm588$ \\ 
6 & $110.26\pm0.12$ & $0.31\pm0.16$ & $0.57\pm0.32$ & $114.0\pm77.3$ & $8\pm2$ & $13\pm9$ & $227\pm154$ \\ 
7 & $91.28\pm0.15$ & $0.88\pm0.18$ & $1.85\pm0.53$ & $1184.3\pm418.4$ & $15\pm2$ & $142\pm50$ & $2368\pm836$ \\ 
8 & $98.67\pm0.12$ & $0.34\pm0.14$ & $0.67\pm0.31$ & $184.7\pm92.8$ & $10\pm1$ & $22\pm11$ & $369\pm185$ \\ 
\noalign{\vspace{3pt}}\hline \noalign{\vspace*{0.8cm}}
\multicolumn{8}{c}{(b) \stwo}  \\
\noalign{\vspace{5pt}}\hline\hline\noalign{\vspace{5pt}}
\multirow{2}{*}{\# \hspace{5pt}} &
\multirow{2}{*}{\hs $\vlsr$ \hs} & 
\multirow{2}{*}{\hs FWHM \hs} & 
\multirow{2}{*}{\hspace{20pt} $F_\mathrm{CO}$\hspace{20pt}} & 
\multirow{2}{*}{\hspace{25pt} $L_\mathrm{CO}$\hspace{25pt}} & 
\multirow{2}{*}{\hs $R_\mathrm{c}$\hs} & 
\multicolumn{2}{c}{\hspace{30pt}$M_\mathrm{mol}$ ($\mo$) \hspace{30pt} } \\ \cmidrule{7-8}
    & (km/s) & (km/s) & (K km/s) & (K km/s pc$^2$) & (pc) & $\alpha_\mathrm{thin}$ & $\alpha_\mathrm{thick}$ \\
\noalign{\vspace{5pt}}\hline\noalign{\vspace{5pt}}
1 & $135.34\pm0.14$ & $0.66\pm0.18$ & $1.15\pm0.35$ & $372.1\pm141.7$ & $11\pm2$ & $44\pm17$ & $744\pm283$ \\ 
2 & $135.08\pm0.18$ & $1.56\pm0.19$ & $2.56\pm0.52$ & $1175.1\pm296.6$ & $13\pm1$ & $141\pm35$ & $2350\pm593$ \\ 
3 & $134.41\pm0.16$ & $0.99\pm0.18$ & $1.71\pm0.44$ & $585.1\pm189.8$ & $11\pm2$ & $70\pm22$ & $1170\pm379$ \\ 
4 & $130.96\pm0.13$ & $0.63\pm0.15$ & $0.97\pm0.28$ & $404.3\pm135.6$ & $12\pm1$ & $48\pm16$ & $808\pm271$ \\ 
5 & $131.50\pm0.14$ & $0.56\pm0.16$ & $0.91\pm0.34$ & $294.4\pm121.9$ & $11\pm1$ & $35\pm14$ & $588\pm243$ \\ 
6 & $135.13\pm0.11$ & $1.03\pm0.14$ & $1.34\pm0.34$ & $388.9\pm120.4$ & $10\pm1$ & $46\pm14$ & $777\pm240$ \\ 
7 & $132.09\pm0.14$ & $0.82\pm0.13$ & $0.63\pm0.27$ & $152.7\pm72.8$ & $9\pm1$ & $18\pm8$ & $305\pm145$ \\ 
8 & $123.56\pm0.18$ & $0.25\pm0.19$ & $0.54\pm0.42$ & $194.1\pm157.0$ & $11\pm2$ & $23\pm18$ & $388\pm313$ \\ 
9 & $123.92\pm0.19$ & $0.33\pm0.21$ & $0.85\pm0.55$ & $337.8\pm238.7$ & $12\pm2$ & $40\pm28$ & $675\pm477$ \\ 
\noalign{\vspace{3pt}}\hline\noalign{\vspace*{0.5cm}}
\end{tabular}
\end{table*}

\subsection{Molecular gas mass}

We used CO-line luminosities and a CO-to-H$_2$ conversion factor $\alphaco$ to calculate the molecular gas mass $M_\mathrm{mol}$ of detected clumps, i.e.\  $M_\mathrm{mol} = \alphaco\colum$ \citep[see][for a comprehensive review]{Bolatto+13}.
We converted from the integrated intensity $\coint$ to CO luminosity $\colum$ as \citep[e.g.,][]{Solomon+97}:

\begin{equation}
\frac{\colum}{(\mathrm{K} \,\kms \, \mathrm{pc}^2)} = 23.5 \, \frac{\coint}{(\mathrm{K} \,\kms)} \frac{\Omega_\mathrm{s}}{(\mathrm{arcs})^2} \left(\frac{D}{\mathrm{Mpc}}\right)^2 
\end{equation}

\noindent where $\Omega_\mathrm{s}$ is the area covered by a clump on the sky and we use a distance $D=0.063$ Mpc.
CO luminosity are listed in column 5 of \autoref{tab:prop}.
The values of the CO-to-H$_2$ conversion factors  have been extensively studied in self-gravitating molecular clouds in the Milky Way \citep[$\alphaco \simeq 4-5 \, \mo \, (\mathrm{K} \, \kms \, \mathrm{pc^2})^{-1}$, e.g.,][]{Heyer+09} and in low-metallicity environments like the SMC \citep[$\alphaco \simeq 15-17 \, \mo \, (\mathrm{K} \, \kms \, \mathrm{pc^2})^{-1}$, e.g.,][]{Jameson+18}.
However, molecular gas in outflows is expected to have very different physical conditions, in terms of density, temperature and pressure, and robust conversion factors have not been established yet. 
Some clues that molecular gas in starburst-driven outflows may be in a lower density and optical-depth state than regular molecular clouds come from CO observations of nearby galaxies M82 \citep[e.g.,][]{Weiss+05} and NGC253 \cite[e.g.,][]{Walter+17}.
A few studies on starburst-driven \citep[e.g.][]{Leroy+15} and AGN-driven outflows \citep[e.g.,][]{Cicone+18} suggest $\alphaco\sim1-2 \, \alphacoun$, lower than the typical values in the Milky Way.

To partially overcome the uncertainties on the CO-to-H$_2$ conversion, we adopt two representative conversion factors. 
Firstly, we calculate lower limits to the molecular gas mass assuming that CO gas is optically thin \citep[see e.g.,][]{Bolatto+13b}. 
In this case, we calculated conversion factors for the CO $J$=2$\rightarrow$1 transition through eq.\ (18) from \citet{Bolatto+13}, setting an energy of the $J=2$ level of $E_2/k = 16.6 \, \mathrm{K}$ and assuming an excitation temperature of $T_\mathrm{ex} = 30 \, \mathrm{K}$ \citep[e.g.,][]{Goldsmith13}.
We obtained a $\alpha_\mathrm{thin} \simeq 0.12 \, \alphacoun$, about two orders of magnitude lower than typical conversion factors in the SMC.
Secondly, we applied the conversion factor $\alpha_\mathrm{thick} \simeq 2 \, \alphacoun$ estimated in local outflows \citep[e.g.,][]{Leroy+15,Cicone+18}, under the assumption that $^{12}$CO(2$\rightarrow$1) is optically thick and thermalized with $^{12}$CO(1$\rightarrow$0) \citep[e.g.,][]{Rubio+93}.
These two conversion factors should give us a plausible range of mass for the molecular gas entrained in the SMC outflow.

The last two columns of Table \ref{tab:prop} list the molecular gas masses calculated through $\alpha_\mathrm{thin}$ and $\alpha_\mathrm{thick}$. 
Typical masses in the optically-thin case are of some tens of $\mo$, while they span between hundreds and a few thousands of $\mo$ in the optically-thick regime.
The total molecular gas mass in the \sone\ field is $\sim500 \, \mo$ and $\sim8\times10^3 \, \mo$ in the optically-thin and -thick cases, $\sim420 \, \mo$ and $\sim7\times10^3 \, \mo$ for the \stwo\ field.
We stress again that masses calculated through $\alpha_\mathrm{thin}$ represent lower limits. 
Considering that it is unlikely that the CO line is completely optically-thin, we can conclude that the mass of molecular gas in both fields is sensibly in the range $10^3-10^4 \, \mo$.

The low metallicity of the SMC could imply larger $\alphaco$ (thus larger molecular gas masses) than the $\alpha_\mathrm{thick}$ estimated in local outflows and assumed so far. 
As a sanity check, we calculated the masses that our CO clouds would have in case of virial equilibrium \citep[see][]{Bolatto+13}, which therefore may represent upper limits to the molecular gas mass for non self-gravitating clouds entrained in an outflow. 
The optically-thick masses calculated with $\alpha_\mathrm{thick}$ are consistent with the virial masses  for spherical clouds with density $\rho(r)\propto r^{-2}$ and about $\sim30\%$  lower than virial masses for clouds with $\rho(r)\propto r^{-1}$. 
This suggests that our molecular masses are not significantly underestimated and that the $\alphaco$ for these CO clumps is likely closer to the 2 $\alphacoun$ estimated in local outflows rather than to the $\sim 16 \, \alphacoun$ found in the main body of the SMC.

\section{Discussion}
\label{sec:discussion}
\citetalias{McG+18} discusses that the most likely interpretation for these \hi\ clouds is a star-formation-driven outflow.
The anomalous kinematics alone would not provide a conclusive argument on the origin of this gas. 
For example, gas accreting onto the galaxy from the surrounding environment \citep{Sancisi+08} and gas stripped because of tidal interactions or because of ram-pressure due to the motion of the SMC through the Milky Way halo \citep{Gunn&Gott72} could easily present peculiar kinematics.
However, a quite compelling evidence in support of the outflow interpretation is the association of these \hi\ features with H$\upalpha$ shells arising from the star-forming regions in the SMC \citepalias[see][for details]{McG+18}.
The detection of CO in both fields studied in this work provides a further support to the outflow scenario: molecular gas is not observed in \hi\ clouds associated with extra-galactic accreting gas \citep[like High-Velocity Clouds, e.g.,][]{DZ+07} and the thermal pressure in a Milky-Way-like halo does not seem high enough to confine stripped gas to the high densities typical of molecular material \citep[e.g., see discussion in][]{Tonnesen&Bryan12}.
However, because we can not constrain the real geometry of the system, we do not know whether the gas is escaping the galaxy or if it is falling back through a galactic fountain mechanism \citep{Fraternali17}. 
\citetalias{McG+18} pointed out that the observed velocity in these clouds is likely large enough to escape the shallow gravitational potential of the SMC.

How hot winds can accelerate cold gas to the observed velocities is still a debated question. 
Several recent simulations of individual cold clouds entrained in a hot, supersonic flow showed that a cloud is easily shredded during the acceleration process on short timescales \citep[e.g.,][]{Scannapieco&Bruggen15,Gronke+18,Sparre+19}.
Although these simulations do not trace the cold molecular phase, the general expectation is of a head-tail cloud morphology, with the densest and coldest core in the trailing head of the cloud and the shredded warmer envelope in the leading tail. 
Our CO and \hi\ observations are qualitatively in agreement with this scenario: if our clouds were launched from the closest prominent star-formation region in the SMC, i.e.\ NGC 371/395, located south-west of \sone\ and south-south-west of \stwo\ (see red box in \autoref{fig:largemap}), the densest CO clumps in our data (4, 5 and 6 in \sone, 1 and 2 in \stwo) would represent the compact core of the initial cloud following the fragmented mixture of molecular and neutral gas.
In this scenario, the spatial distributions of neutral and molecular gas and the observed offsets are signatures of the cloud disruption and of the interaction between cold gas and the hot flow.
A puzzling aspect that emerges from our observations and seems to hinder the entrainment interpretation is the very narrow linewidths observed in the molecular gas ($\sim 1 \, \kms$): velocity dispersions are expected to increase while cold gas is shocked and shredded by the hot flow \citep{Banda+19}. 
New simulations aimed to trace properly the molecular phase of the outflow are needed to address this observational evidence.
The small linewidths may imply that we are only observing the innermost region of a molecular cloud which is not (yet) directly interacting with the wind.


The total mass of molecular gas in both our targets is $10^3-10^4 \, \mo$.
The amount of neutral gas (H+He) estimated from the \hi\ ASKAP observations in the same regions of our APEX fields is $4.0\times10^4 \, \mo$ for \sone\ and $3.3\times10^4 \, \mo$ for \stwo. 
This leads to molecular gas fractions $f_\mathrm{mol} =  M_\mathrm{mol} / ( M_\mathrm{\hi} + M_\mathrm{mol}) \simeq 0.03 - 0.30$, i.e.\ the contribution in mass of molecular gas to the total outflowing material ranges from a few percent up to $\sim30\%$.
\citetalias{McG+18} estimated an outflow rate in \hi\ of $\dot{M}_\mathrm{\hi} \simeq 0.2 - 1 \, \moyr$ ($\sim 0.3 - 1.4 \, \moyr$ including Helium), assuming that the outflowing gas originated from the most recent burst of star formation in the SMC, between 25 and 60 Myr ago \citep[e.g.,][]{Rubele+15}.
If the molecular gas fractions derived in this work hold for all the anomalous \hi\ features found by \citetalias{McG+18}, the cold gas (neutral + molecular) outflow rate adds up to a value $\dot{M}_\mathrm{gas} \sim 0.3 - 1.8 \, \moyr$.
If only 40\% of the ejecta have velocities exceeding the escape velocity from the galaxy \citepalias[see][]{McG+18}, the SMC may lose its entire cold gas reservoir, i.e.\ $\sim6\times10^8 \, \mo$ of neutral gas \citep[e.g.,][]{Stanimirovic+99} and $\sim3\times10^7 \, \mo$ of molecular gas \citep[e.g.,][]{Rubio+93}, on timescales of $t_\mathrm{dep} = 0.9-3$ Gyr.
We speculate that this large amount of cold gas expelled from the SMC may feed and enrich the Magellanic Stream and the circum-galactic medium of the Milky Way.

Finally, we can estimate a cold mass loading factor of $\beta \equiv \dot{M}_\mathrm{gas} / \mathrm{SFR} \simeq 3 - 12$, where $\mathrm{SFR} \simeq 0.15 \, \moyr$ is the average star-formation rate during the period $25-60$ Myr ago \citep[e.g.,][]{Rubele+15}.
Comparing these numbers to simulations of SN-driven winds in dwarf galaxies is not straightforward because
mass loading factors are a function of the distance from the wind launching region. 
In addition, most simulations do not appropriately trace neutral and molecular gas phases and quoted loading factors often include the hot ionized phase.
If we assume that our clouds started their journey from the NGC371 region, we have angular separations of $\sim 0.8\de$ and $\sim1\de$ for \sone\ and \stwo, respectively, corresponding to projected distances of $\sim 0.9$ kpc and $\sim1.1$ kpc. 
These are lower limits to the actual distances, which are unconstrainable due to the unknown geometry of the ejected gas. 
Since most of the \hi\ outflowing clouds in \citetalias{McG+18} lie at projected distances $\gtrsim 1$ kpc from the closest star formation regions, our estimated loading factor is measured at $R> 1$ kpc.
The general theoretical expectation is that dwarf galaxies drive out gas at a rate much higher than their SFR, i.e.\ $\beta\gg1$. 
For examples, cosmological simulations by \cite{Hopkins+12} and \citet{Muratov+15} returned loading factors (including the hot phase) at $R>500$ pc of $\beta\sim8-20$ for SMC-like dwarf galaxies. 
Recent high-resolution simulations of SN-driven winds in isolated dwarf galaxies by \citet{Hu19} found $\beta\sim1-10$ for the warm gas component ($T<3\times10^4$ K) at $R>1$ kpc.
Although the exact values of loading factors depend also on the initial conditions for the gas (e.g.\ density, metallicity, thickness of the disk) and on the prescriptions adopted for star formation and SN feedback, we can conclude that the cold mass loading factor $\beta = 3-12$ estimated in this work is broadly consistent with those found in simulations.

\section{Conclusions}
\label{sec:conclusions}

This Letter presented the first study of molecular gas associated with the outflow of the Small Magellanic Cloud through the detection of the \co\ emission line with the APEX telescope. 
Thanks to the relatively small distance of the SMC, we spatially resolved individual knots of CO emission in two outflowing \hi\ clouds.
CO clumps are compact ($\sim 10$ pc), cold and non-turbulent, with typical integrated line broadening of less than 1 $\kms$.
Most CO clumps do not overlap but they are slightly offset from the densest \hi\ regions, in qualitative agreement with expectations from simulations of cold clouds entrained in a hot wind. 
We estimated molecular gas masses in optically thin ($\alphaco = 0.12 \, \alphacoun$) and optically thick outflow ($\alphaco = 2 \, \alphacoun$) regimes and we ended up with a range of masses $M_\mathrm{mol} = 10^3-10^4 \, \mo$ for both the observed fields, with molecular to total cold gas fractions between 0.03 and 0.30.
Assuming an outflow interpretation, we estimated a total outflow rate of $0.3-1.8 \, \mo$ for the cold gas component, implying that the SMC may expel the majority of its present-day cold gas reservoir in a few Gyr. 
The estimated cold-mass loading-factor $\beta \simeq 3-12$ at $R>1$ kpc is overall consistent with simulations of SN-driven winds in dwarf galaxies. 


\acknowledgments
E.D.T.\ thanks Alberto Bolatto for his useful comments and P.~F.~Winkler for providing MCELS data.
E.D.T.\ and N.M.-G.\ acknowledge the support of the Australian Research Council through grant DP160100723.
M.R. wishes to acknowledge support from CONICYT(CHILE) through FONDECYT grant No1140839 and partial support from CONICYT project Basal AFB-170002
APEX is a collaboration between the Max-Planck-Institut fur Radioastronomie, the European Southern Observatory, and the Onsala Space Observatory.

\vspace*{0.5cm}
\facilities{APEX(PI230), ASKAP}

\software{CLASS/GILDAS \citep{Gildas+13},  $^\mathrm{3D}$\textsc{Barolo} \citep{DiTeodoro&Fraternali15} }

\bibliography{bib_Clouds}{}

\begin{thebibliography}{}
\expandafter\ifx\csname natexlab\endcsname\relax\def\natexlab#1{#1}\fi

\bibitem[{{Armillotta} {et~al.}(2019){Armillotta}, {Krumholz}, {Di Teodoro}, \&
  {McClure-Griffiths}}]{Armillotta+19}
{Armillotta}, L., {Krumholz}, M.~R., {Di Teodoro}, E.~M., \&
  {McClure-Griffiths}, N.~M. 2019, \mnras, 2479

\bibitem[{{Arribas} {et~al.}(2014){Arribas}, {Colina}, {Bellocchi}, {Maiolino},
  \& {Villar-Mart{\'\i}n}}]{Arribas+14}
{Arribas}, S., {Colina}, L., {Bellocchi}, E., {Maiolino}, R., \&
  {Villar-Mart{\'\i}n}, M. 2014, Astronomy and Astrophysics, 568, A14

\bibitem[{{Banda-Barrag{\'a}n} {et~al.}(2019){Banda-Barrag{\'a}n}, {Zertuche},
  {Federrath}, {Garc{\'\i}a Del Valle}, {Br{\"u}ggen}, \& {Wagner}}]{Banda+19}
{Banda-Barrag{\'a}n}, W.~E., {Zertuche}, F.~J., {Federrath}, C., {et~al.} 2019,
  \mnras, 486, 4526

\bibitem[{{Bolatto} {et~al.}(2013{\natexlab{a}}){Bolatto}, {Wolfire}, \&
  {Leroy}}]{Bolatto+13}
{Bolatto}, A.~D., {Wolfire}, M., \& {Leroy}, A.~K. 2013{\natexlab{a}}, \araa,
  51, 207

\bibitem[{{Bolatto} {et~al.}(2013{\natexlab{b}}){Bolatto}, {Warren}, {Leroy},
  {Walter}, {Veilleux}, {Ostriker}, {Ott}, {Zwaan}, {Fisher}, {Weiss},
  {Rosolowsky}, \& {Hodge}}]{Bolatto+13b}
{Bolatto}, A.~D., {Warren}, S.~R., {Leroy}, A.~K., {et~al.} 2013{\natexlab{b}},
  \nat, 499, 450

\bibitem[{{Cicone} {et~al.}(2018){Cicone}, {Severgnini}, {Papadopoulos},
  {Maiolino}, {Feruglio}, {Treister}, {Privon}, {Zhang}, {Della Ceca}, {Fiore},
  {Schawinski}, \& {Wagg}}]{Cicone+18}
{Cicone}, C., {Severgnini}, P., {Papadopoulos}, P.~P., {et~al.} 2018, \apj,
  863, 143

\bibitem[{{Dessauges-Zavadsky} {et~al.}(2007){Dessauges-Zavadsky}, {Combes}, \&
  {Pfenniger}}]{DZ+07}
{Dessauges-Zavadsky}, M., {Combes}, F., \& {Pfenniger}, D. 2007, \aap, 473, 863

\bibitem[{{Di Teodoro} \& {Fraternali}(2015)}]{DiTeodoro&Fraternali15}
{Di Teodoro}, E.~M., \& {Fraternali}, F. 2015, \mnras, 451, 3021

\bibitem[{{Fraternali}(2017)}]{Fraternali17}
{Fraternali}, F. 2017, in Astrophysics and Space Science Library, Vol. 430, Gas
  Accretion onto Galaxies, ed. A.~{Fox} \& R.~{Dav{\'e}}, 323

\bibitem[{{Gildas Team}(2013)}]{Gildas+13}
{Gildas Team}. 2013, {GILDAS: Grenoble Image and Line Data Analysis Software},
  Astrophysics Source Code Library, , , ascl:1305.010

\bibitem[{{Goldsmith}(2013)}]{Goldsmith13}
{Goldsmith}, P.~F. 2013, \apj, 774, 134

\bibitem[{{Gronke} \& {Oh}(2018)}]{Gronke+18}
{Gronke}, M., \& {Oh}, S.~P. 2018, \mnras, 480, L111

\bibitem[{{Gunn} \& {Gott}(1972)}]{Gunn&Gott72}
{Gunn}, J.~E., \& {Gott}, J.~Richard, I. 1972, \apj, 176, 1

\bibitem[{{G{\"u}sten} {et~al.}(2006){G{\"u}sten}, {Nyman}, {Schilke},
  {Menten}, {Cesarsky}, \& {Booth}}]{Gusten+06}
{G{\"u}sten}, R., {Nyman}, L.~{\r{A}}., {Schilke}, P., {et~al.} 2006, \aap,
  454, L13

\bibitem[{{Heyer} {et~al.}(2009){Heyer}, {Krawczyk}, {Duval}, \&
  {Jackson}}]{Heyer+09}
{Heyer}, M., {Krawczyk}, C., {Duval}, J., \& {Jackson}, J.~M. 2009, \apj, 699,
  1092

\bibitem[{{Hopkins} {et~al.}(2013){Hopkins}, {Kere{\v{s}}}, {Murray},
  {Hernquist}, {Narayanan}, \& {Hayward}}]{Hopkins+13}
{Hopkins}, P.~F., {Kere{\v{s}}}, D., {Murray}, N., {et~al.} 2013, \mnras, 433,
  78

\bibitem[{{Hopkins} {et~al.}(2012){Hopkins}, {Quataert}, \&
  {Murray}}]{Hopkins+12}
{Hopkins}, P.~F., {Quataert}, E., \& {Murray}, N. 2012, \mnras, 421, 3522

\bibitem[{{Hu}(2019)}]{Hu19}
{Hu}, C.-Y. 2019, \mnras, 483, 3363

\bibitem[{{Jameson} {et~al.}(2018){Jameson}, {Bolatto}, {Wolfire}, {Warren},
  {Herrera-Camus}, {Croxall}, {Pellegrini}, {Smith}, {Rubio}, {Indebetouw},
  {Israel}, {Meixner}, {Roman-Duval}, {van Loon}, {Muller}, {Verdugo},
  {Zinnecker}, \& {Okada}}]{Jameson+18}
{Jameson}, K.~E., {Bolatto}, A.~D., {Wolfire}, M., {et~al.} 2018, \apj, 853,
  111

\bibitem[{{Kim} \& {Ostriker}(2018)}]{Kim&Ostriker18}
{Kim}, C.-G., \& {Ostriker}, E.~C. 2018, \apj, 853, 173

\bibitem[{{Klein} {et~al.}(2012){Klein}, {Hochg{\"u}rtel}, {Kr{\"a}mer},
  {Bell}, {Meyer}, \& {G{\"u}sten}}]{Klein+12}
{Klein}, B., {Hochg{\"u}rtel}, S., {Kr{\"a}mer}, I., {et~al.} 2012, \aap, 542,
  L3

\bibitem[{{Leroy} {et~al.}(2015){Leroy}, {Walter}, {Martini}, {Roussel},
  {Sandstrom}, {Ott}, {Weiss}, {Bolatto}, {Schuster}, \&
  {Dessauges-Zavadsky}}]{Leroy+15}
{Leroy}, A.~K., {Walter}, F., {Martini}, P., {et~al.} 2015, \apj, 814, 83

\bibitem[{{Mart{\'\i}n-Fern{\'a}ndez}
  {et~al.}(2016){Mart{\'\i}n-Fern{\'a}ndez}, {Jim{\'e}nez-Vicente}, {Zurita},
  {Mediavilla}, \& {Castillo-Morales}}]{Martin+16}
{Mart{\'\i}n-Fern{\'a}ndez}, P., {Jim{\'e}nez-Vicente}, J., {Zurita}, A.,
  {Mediavilla}, E., \& {Castillo-Morales}, {\'A}. 2016, \mnras, 461, 6

\bibitem[{{McClure-Griffiths} {et~al.}(2018){McClure-Griffiths}, {D{\'e}nes},
  {Dickey}, {Stanimirovi{\'c}}, {Staveley-Smith}, {Jameson}, {Di Teodoro},
  {Allison}, {Collier}, \& {Chippendale}}]{McG+18}
{McClure-Griffiths}, N.~M., {D{\'e}nes}, H., {Dickey}, J.~M., {et~al.} 2018,
  Nature Astronomy, 2, 901

\bibitem[{{Muratov} {et~al.}(2015){Muratov}, {Kere{\v{s}}},
  {Faucher-Gigu{\`e}re}, {Hopkins}, {Quataert}, \& {Murray}}]{Muratov+15}
{Muratov}, A.~L., {Kere{\v{s}}}, D., {Faucher-Gigu{\`e}re}, C.-A., {et~al.}
  2015, \mnras, 454, 2691

\bibitem[{{Rubele} {et~al.}(2015){Rubele}, {Girardi}, {Kerber}, {Cioni},
  {Piatti}, {Zaggia}, {Bekki}, {Bressan}, {Clementini}, {de Grijs}, {Emerson},
  {Groenewegen}, {Ivanov}, {Marconi}, {Marigo}, {Moretti}, {Ripepi},
  {Subramanian}, {Tatton}, \& {van Loon}}]{Rubele+15}
{Rubele}, S., {Girardi}, L., {Kerber}, L., {et~al.} 2015, \mnras, 449, 639

\bibitem[{{Rubio} {et~al.}(1993){Rubio}, {Lequeux}, \& {Boulanger}}]{Rubio+93}
{Rubio}, M., {Lequeux}, J., \& {Boulanger}, F. 1993, \aap, 271, 9

\bibitem[{{Sancisi} {et~al.}(2008){Sancisi}, {Fraternali}, {Oosterloo}, \& {van
  der Hulst}}]{Sancisi+08}
{Sancisi}, R., {Fraternali}, F., {Oosterloo}, T., \& {van der Hulst}, T. 2008,
  \aapr, 15, 189

\bibitem[{{Scannapieco} \& {Br{\"u}ggen}(2015)}]{Scannapieco&Bruggen15}
{Scannapieco}, E., \& {Br{\"u}ggen}, M. 2015, \apj, 805, 158

\bibitem[{{Solomon} {et~al.}(1997){Solomon}, {Downes}, {Radford}, \&
  {Barrett}}]{Solomon+97}
{Solomon}, P.~M., {Downes}, D., {Radford}, S.~J.~E., \& {Barrett}, J.~W. 1997,
  \apj, 478, 144

\bibitem[{{Sparre} {et~al.}(2019){Sparre}, {Pfrommer}, \&
  {Vogelsberger}}]{Sparre+19}
{Sparre}, M., {Pfrommer}, C., \& {Vogelsberger}, M. 2019, \mnras, 482, 5401

\bibitem[{{Stanimirovic} {et~al.}(1999){Stanimirovic}, {Staveley-Smith},
  {Dickey}, {Sault}, \& {Snowden}}]{Stanimirovic+99}
{Stanimirovic}, S., {Staveley-Smith}, L., {Dickey}, J.~M., {Sault}, R.~J., \&
  {Snowden}, S.~L. 1999, \mnras, 302, 417

\bibitem[{{Tanner} {et~al.}(2016){Tanner}, {Cecil}, \& {Heitsch}}]{Tanner+16}
{Tanner}, R., {Cecil}, G., \& {Heitsch}, F. 2016, \apj, 821, 7

\bibitem[{{Tonnesen} \& {Bryan}(2012)}]{Tonnesen&Bryan12}
{Tonnesen}, S., \& {Bryan}, G.~L. 2012, \mnras, 422, 1609

\bibitem[{{Veilleux} {et~al.}(2005){Veilleux}, {Cecil}, \&
  {Bland-Hawthorn}}]{Veilleux+05}
{Veilleux}, S., {Cecil}, G., \& {Bland-Hawthorn}, J. 2005, \araa, 43, 769

\bibitem[{{Walter} {et~al.}(2017){Walter}, {Bolatto}, {Leroy}, {Veilleux},
  {Warren}, {Hodge}, {Levy}, {Meier}, {Ostriker}, {Ott}, {Rosolowsky},
  {Scoville}, {Weiss}, {Zschaechner}, \& {Zwaan}}]{Walter+17}
{Walter}, F., {Bolatto}, A.~D., {Leroy}, A.~K., {et~al.} 2017, \apj, 835, 265

\bibitem[{{Wei{\ss}} {et~al.}(2005){Wei{\ss}}, {Walter}, \&
  {Scoville}}]{Weiss+05}
{Wei{\ss}}, A., {Walter}, F., \& {Scoville}, N.~Z. 2005, \aap, 438, 533

\bibitem[{{Winkler} {et~al.}(2015){Winkler}, {Smith}, {Points}, \& {MCELS
  Team}}]{Winkler+15}
{Winkler}, P.~F., {Smith}, R.~C., {Points}, S.~D., \& {MCELS Team}. 2015, in
  Astronomical Society of the Pacific Conference Series, Vol. 491, Fifty Years
  of Wide Field Studies in the Southern Hemisphere: Resolved Stellar
  Populations of the Galactic Bulge and Magellanic Clouds, ed. S.~{Points} \&
  A.~{Kunder}, 343

\bibitem[{{Zhang}(2018)}]{Zhang18}
{Zhang}, D. 2018, Galaxies, 6, 114

\end{thebibliography}
\bibliographystyle{aasjournal}

\appendix
\section{Observations and data reduction}
\label{app}
Sub-millimeter observations were made using the 12m APEX telescope in June 2019 (ESO project ID 0103.B-0120A; P.I. Di Teodoro E.~M.). The weather conditions were stable and dry, with a precipitable water vapour $\mathrm{PWV}=0.6-1.5$ mm. The PI230 heterodyne receiver was tuned to map the \co\ emission line at 230.538 GHz. The 4th-generation fast-Fourier transform spectrometer \citep[dFFTS4G,][]{Klein+12} backend connected to the receiver provided a bandwidth of 8 GHz with a spectral resolution of 61 kHz, corresponding to a velocity resolution of about 0.08 $\kms$ at 230 GHz. At this frequency, the beam size is $\theta=26.3''$ (FWHM), the main-beam efficiency is $\eta_\mathrm{mb}\simeq0.72$ and the Jy/K conversion factor is $40\pm3$\footnote{\href{http://www.apex-telescope.org/telescope/efficiency/}{http://www.apex-telescope.org/telescope/efficiency}}. We observed our targets in on-the-fly total-power mode, sampling every 9$''$ with integration time of 1 second. For \sone, we mapped a $10' \times 10'$ region centered at $(\alpha,\delta)_\mathrm{J2000} = (\mathrm{01h08m58.0s},-71^\circ 22' 39'')$; for \stwo, we mapped a $7' \times 10'$ region centered at $(\alpha,\delta)_\mathrm{J2000} = (\mathrm{01h 05m 31.0s},-71^\circ 04' 26'')$. Observed regions are shown as red boxes in \autoref{fig:clouds}. Integration times were 25h and 17h for \sone\ and \stwo\, respectively, including calibrations and overheads.

Data reduction was performed using the Continuum and Line Analysis Single-dish Software \citep[CLASS,][]{Gildas+13}.
A first-order baseline was subtracted from calibrated spectra by interpolating the channels outside the velocity windows where we expected to see the emission based on the \hi\ observations. We resampled the spectra with a 0.25 $\kms$ channel width and mapped them onto a grid with pixel sizes of $9''$. The root mean square (rms) noise in our final datacubes is 37 mK and 32 mK in a 0.25 $\kms$ channel for \sone\ and \stwo, respectively.

\end{document}